\documentclass[a4paper,11pt]{article}
\usepackage{jheppub} % for details on the use of the package, please see the JINST-author-manual
\usepackage{subfigure}
\newcommand{\mm}{M_{\rm{miss}}}
\newcommand{\Lc}{\Lambda_c^{+}}
\newcommand{\ALc}{\bar{\Lambda}_c^{-}}
\newcommand{\Lcnpieta}{\Lambda_c^{+}\to n\pi^+\eta}
\newcommand{\Lclmdpieta}{\Lambda_c^{+}\to\Lambda\pi^+\eta}
\newcommand{\Lcsgmpieta}{\Lambda_c^{+}\to\Sigma^0\pi^+\eta}
\newcommand{\Lcxpieta}{\Lambda_c^{+}\to X\pi^+\eta\,(X=n,\,\Lambda,\,\Sigma^{0})}
\newcommand{\etagg}{\eta\to\gamma\gamma}
\newcommand{\etatpi}{\eta\to\pi^+\pi^-\pi^0}

\def\gev{\ifmmode {\mathrm{\ Ge\kern -0.1em V}}\else
                   \textrm{Ge\kern -0.1em V}\fi}%
\def\mev{\ifmmode {\mathrm{\ Me\kern -0.1em V}}\else
                   \textrm{Me\kern -0.1em V}\fi}%
\def\ifb{\mbox{fb$^{-1}$}}

\newcommand{\BESIIIorcid}[1]{\href{https://orcid.org/#1}{\hspace*{0.1em}\raisebox{-0.45ex}{\includegraphics[width=1em]{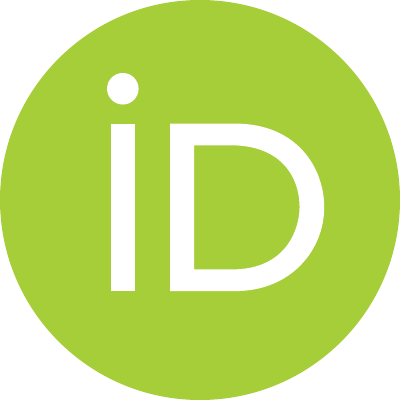}}}}

\arxivnumber{2603.28232} % if you have one
\title{\boldmath Observation of $\Lambda^+_c\to n\pi^+\eta$ and search for $\Lambda^+_c\to na_0(980)^+$}

\collaborationImg{\includegraphics[height=30mm,angle=90]{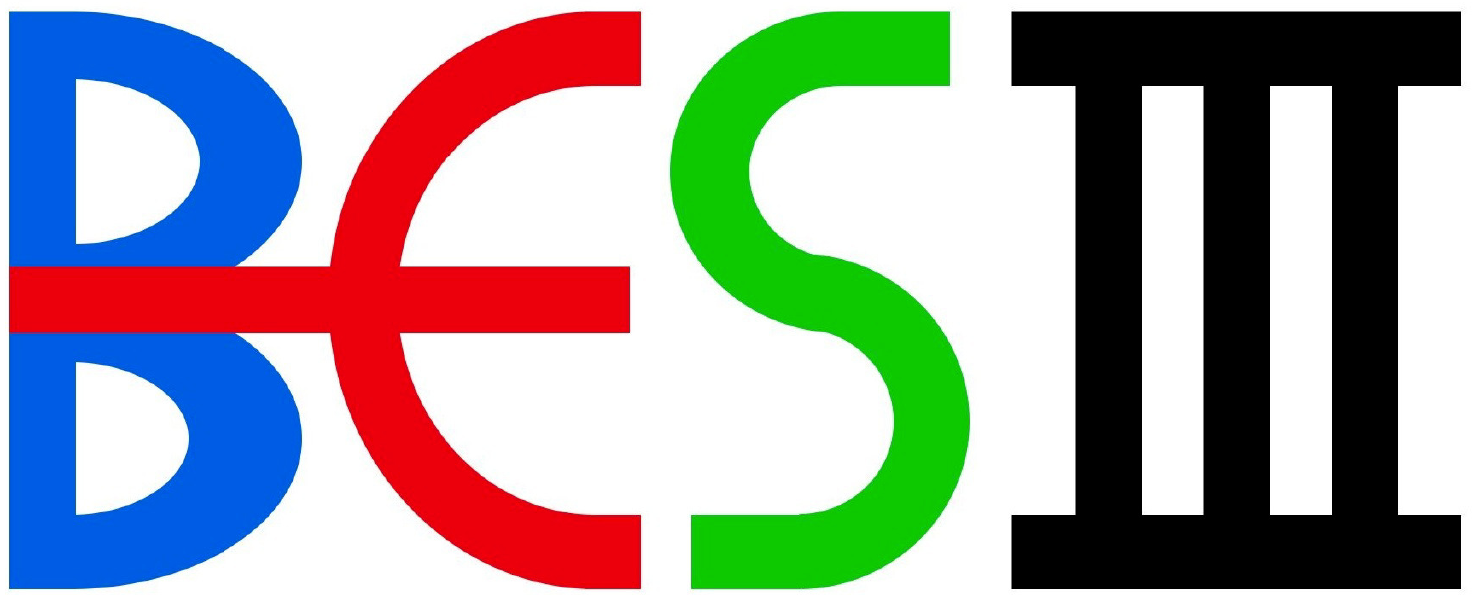}}
\collaboration{The BESIII collaboration}
\emailAdd{besiii-publications@ihep.ac.cn}

\abstract{By analysing 6.1 ${\rm fb}^{-1}$ of data collected at centre-of-mass energies between $\sqrt{s}=4.600$ and 4.843 $\rm GeV$ with the BESIII detector at the BEPCII collider, we observe the decay $\Lambda_c^+\to n\pi^+\eta$ for the first time with a statistical significance of $9.5\sigma$. The ratio of branching fractions $\mathcal{B}(\Lambda_c^+\to n\pi^+\eta)/\mathcal{B}(\Lambda_c^+\to \Lambda\pi^+\eta)$ is measured to be $0.155\pm0.031_{\rm stat.}\pm0.012_{\rm syst.}$ Taking the world average of $\mathcal{B}(\Lambda_c^+\to \Lambda\pi^+\eta)$ as reference, the absolute branching fraction is calculated to be $\mathcal{B}(\Lambda_c^+\to n\pi^+\eta)=(2.94\pm0.59_{\rm stat.}\pm0.23_{\rm syst.}\pm0.13_{\rm ref.})\times10^{-3}$. The intermediate process $\Lambda_c^+\to na_0(980)^+$ is also searched for in the $\pi^+\eta$ invariant mass spectrum. Since no significant signal is found, the upper limit on $\mathcal{B}(\Lambda_c^+\to na_0(980)^+)\times\mathcal{B}(a_0(980)^+\to\pi^+\eta)$ is set to $8.4\times10^{-4}$ at 90\% confidence level. A sophisticated deep learning approach using a Transformer-based architecture is employed to distinguish signals from prevalent hadronic backgrounds, complemented by thorough validation and systematic uncertainty quantification.}


\begin{document}
\maketitle
\flushbottom

\section{Introduction}\label{sec:intro}
\hspace{1.5em}
The light scalar meson $a_0(980)$ remains enigmatic, with its underlying structure interpreted variously in the literature: as a conventional $q\bar{q}$ state~\cite{Soni:2020sgn,Klempt:2021nuf}, a compact tetraquark configuration~\cite{Achasov:1999wv,Jaffe:2004ph}, or even a quantum superposition of both~\cite{Alexandrou:2017itd}. An alternative hypothesis posits that the $a_0(980)$ may emerge as a dynamically generated threshold resonance~\cite{Weinstein:1982gc,Janssen:1994wn,Oller:1997ti,Sekihara:2014qxa,Ahmed:2020kmp}. Decay channels of the charm baryon $\Lambda_c^+$ provide a unique laboratory for probing the properties of $a_0(980)$. Recently, the BESIII collaboration reported the decay $\Lambda_c^+\to\Lambda\pi^+\eta$~\cite{BESIII:2024mbf}, which revealed a striking anomaly. Through partial-wave analysis (PWA), the intermediate process $\Lambda_c^+\to\Lambda a_0(980)^+$, with $a_0(980)^+\to\pi^+\eta$, was isolated. Notably, the measured branching fraction (BF) of $\Lambda_c^+\to\Lambda a_0(980)^+$ exhibits order-of-magnitude discrepancies with theoretical predictions~\cite{Sharma:2009zze,Yu:2020vlt}. Neither short-distance nor long-distance contributions satisfactorily account for the observed dynamics, potentially hinting at an exotic substructure of the $a_0(980)^+$.

\begin{figure}[htbp]
	\centering
	\subfigure[Factorizable $W$-emission diagram]
	{\label{fig:Diagram_na980_fact}
	\includegraphics[width=0.30\textwidth]{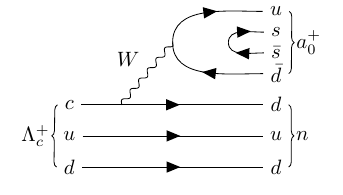}}
	\subfigure[Non-factorizable $W$-emission diagram]
	{\label{fig:Diagram_na980_nonfact1}
    \includegraphics[width=0.30\textwidth]{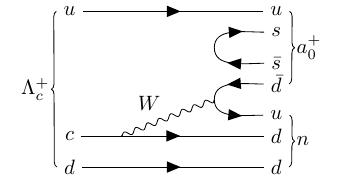}}
    \subfigure[Non-factorizable $W$-exchange diagram]
	{\label{fig:Diagram_na980_nonfact2}
	\includegraphics[width=0.30\textwidth]{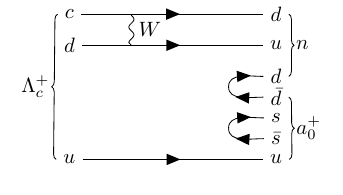}}
	\caption{Topological diagrams for $\Lambda_c^+\to n a_0(980)^+$ assuming $a_0(980)^+$ as a tetraquark state.\label{fig:Diagram_na980_W}}
\end{figure}

Replacing the final-state $\Lambda$ hyperon with a neutron, the singly Cabibbo-suppressed (SCS) decay $\Lambda_c^+\to n a_0(980)^+$ and its three-body counterpart $\Lcnpieta$ offer a unique probe of the $a_0(980)^+$ structure. Theoretical predictions for these decays exhibit significant variations. Under the tetraquark assumption, Ref.~\cite{Sharma:2009zze} calculates the amplitude through both factorizable and non-factorizable diagrams. The former contribution depends on the $a_0(980)^+$ decay constant and $\Lambda_c^+\to n$ transition form factors, while the latter employs pole models with positive-parity intermediate baryons, yielding $\mathcal{B}(\Lambda_c^+\to n a_0(980)^+)=7.1\times10^{-5}$. However, this result carries large uncertainties due to poorly constrained inputs including the light scalar decay constant, $\Lambda_c^+\to n$ form factors, and strong baryon--scalar-meson coupling constants. A more recent calculation based on $SU(3)_f$ flavor symmetry~\cite{Wang:2025uie} adopts a topological diagram amplitude approach, as illustrated in Fig.~\ref{fig:Diagram_na980_W}, predicting $\mathcal{B}(\Lambda_c^+\to na_0(980)^+)=(1.4\pm1.0)\times 10^{-3}$ for the two-quark scenario and $(2.3\pm1.1)\times 10^{-3}$ for the tetraquark scenario, where the substantial uncertainties reflect current theoretical limitations. These order-of-magnitude discrepancies, coupled with strong model dependence, highlight the critical need for direct experimental measurements of $\Lambda_c^+\to na_0(980)^+$ to distinguish between competing quark-configuration hypotheses and improve QCD-based theoretical frameworks.

\begin{figure}[ht]
	\centering
	\subfigure[]
	{\label{fig:Diagram_a2}
	\includegraphics[width=0.31\textwidth]{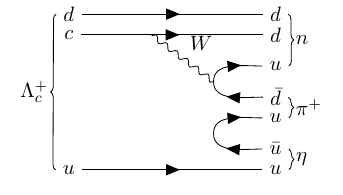}}
    \subfigure[]
	{\label{fig:Diagram_a4_3}
	\includegraphics[width=0.31\textwidth]{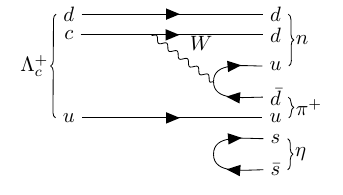}}
    \subfigure[]
	{\label{fig:Diagram_a5}
	\includegraphics[width=0.31\textwidth]{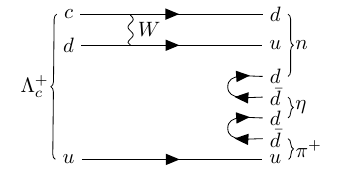}}
    \subfigure[]
	{\label{fig:Diagram_a7}
	\includegraphics[width=0.31\textwidth]{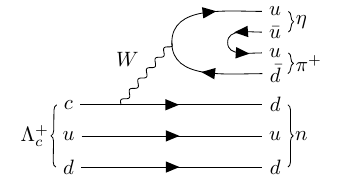}}
    \subfigure[]
	{\label{fig:Diagram_a8}
	\includegraphics[width=0.31\textwidth]{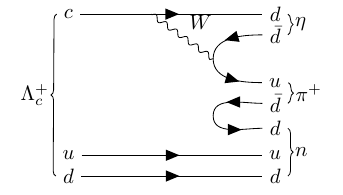}}
	\caption{Topological diagrams for the $\Lambda_c^+\to n\pi^+\eta$ process, where (a--c) receive contributions from $H(\bar{\textbf{6}})$, while (d) and (e) receive contributions from $H(\textbf{15})$.\label{fig:Diagram_npieta_SU3}}
\end{figure}

Although the study of $\Lambda_c^+\to n a_0(980)^+$ is of significant interest, precise characterisation of the three-body decay $\Lcnpieta$ is essential before probing the subprocess. Theoretical studies of such three-body charm baryon decays have been conducted within the $SU(3)_f$ flavor symmetry framework, where pseudoscalar meson-pair final states are typically assumed to be dominated by $S$-wave configurations, with final-state interactions often neglected~\cite{Geng:2018upx,Cen:2019ims,Geng:2024sgq}. A recent work by Geng {\it et al.}~\cite{Geng:2024sgq} fully incorporates contributions from both the color-antisymmetric $\bar{\textbf{6}}$ and color-symmetric $\textbf{15}$ components of the effective Hamiltonian, as shown in Fig.~\ref{fig:Diagram_npieta_SU3}, predicting $\mathcal{B}(\Lambda_c^+\to n\pi^+\eta) = (4.52\pm1.21)\times10^{-3}$ and a decay asymmetry of $0.6^{+0.4}_{-0.5}$. Alternatively, Li {\it et al.}~\cite{Li:2025gvo} investigate $\Lcnpieta$ by explicitly including intermediate states such as $a_0(980)^+$ and $N(1535)^0$, dynamically generated through $S$-wave pseudoscalar-pseudoscalar and pseudoscalar-baryon interactions in a chiral unitary approach. Their analysis suggests a BF ratio of $\mathcal{B}(\Lambda_c^+\to n a_0(980)^+)/\mathcal{B}(\Lcnpieta) \approx 0.313$. Future measurements of $\Lcnpieta$, particularly the isolation of the $\Lambda_c^+\to n a_0(980)^+$ component, would provide invaluable constraints on the dynamics of charm baryon decays and the role of final-state interactions.

On the experimental side, the decay $\Lcnpieta$ has not been investigated previously. Insights from an analogous decay channel, $\Lc\to n\pi^+\pi^0$~\cite{BESIII:2022xne}, indicate substantial background contamination arising from processes with and without $\Lc$ decays. Conventional cut-based analysis methods fail to extract the $\Lcnpieta$ signal above backgrounds with adequate statistical significance. The need for an unprecedentedly powerful signal identification tool has led us to adopt a deep learning approach, in which a deep neural network (DNN) is trained to classify signal and background events. Owing to flexible data representation and modern algorithms, this method has demonstrated remarkable capabilities in uncovering potential relationships and hidden patterns among final-state particles. It has also achieved notable success in previous BESIII studies~\cite{BESIII:2024mgg,BESIII:2024cbr,BESIII:2025biy,BESIII:2026lqo}. 

In this paper, we present the first experimental study of the decay $\Lcnpieta$ by analysing 6.1~$\ifb$ of data collected at centre-of-mass energies between $\sqrt{s}=4.600$ and $4.843~\gev$~\cite{BESIII:2015qfd,BESIII:2015zbz,BESIII:2022dxl,BESIII:2022ulv}, where $\Lambda_c^+$ is dominantly produced via pair production $e^+e^-\to\Lambda_c^+\bar{\Lambda}_c^-$. 

The paper is organised as follows. Section~\ref{sec:detector_and_MC} provides a brief description of the BESIII detector and Monte Carlo (MC) simulations. Section~\ref{sec:BF} details the measurement of the BF for $\Lambda_c^+\to n\pi^+\eta$, encompassing the general analysis strategy, primary event selection, signal-background separation using a deep learning technique, BF extraction, and systematic uncertainty estimation. In Section~\ref{sec:fita0}, we report the search for the intermediate process $\Lambda_c^+\to na_0(980)^+$. A summary and conclusions are given in Section~\ref{sec:summary}. Unless explicitly stated otherwise, charge conjugation is implied throughout this paper.

\section{BESIII detector and Monte Carlo simulation}\label{sec:detector_and_MC}
\hspace{1.5em}
The BESIII detector~\cite{BESIII:2009fln} records symmetric $e^+e^-$ collisions provided by the BEPCII storage ring~\cite{Yu:IPAC2016-TUYA01} in the centre-of-mass energy range from 1.84 to 4.95~GeV, with a peak luminosity of $1.1 \times 10^{33}\;\text{cm}^{-2}\text{s}^{-1}$ achieved at $\sqrt{s} = 3.773\;\text{GeV}$. BESIII has collected large data samples in this energy region~\cite{BESIII:2020nme,Lu:2020imt,Zhang:2022bdc}. The cylindrical core of the BESIII detector covers 93\% of the full solid angle and consists of a helium-based multilayer drift chamber~(MDC), a time-of-flight system~(TOF), and a CsI(Tl) electromagnetic calorimeter~(EMC), which are all enclosed in a superconducting solenoidal magnet providing a 1.0~T magnetic field. The solenoid is supported by an octagonal flux-return yoke with resistive plate counter muon identification modules interleaved with steel. The charged-particle momentum resolution at $1~{\rm GeV}/c$ is $0.5\%$, and the ${\rm d}E/{\rm d}x$ resolution is $6\%$ for electrons from Bhabha scattering. The EMC measures photon energies with a resolution of $2.5\%$ ($5\%$) at $1$~GeV in the barrel (end cap) region. The time resolution in the plastic scintillator TOF barrel region is 68~ps, while that in the end cap region was 110~ps. The end cap TOF system was upgraded in 2015 using multigap resistive plate chamber technology, providing a time resolution of
60~ps, which benefits 90\% of the data used in this analysis~\cite{Li:2017jpg,Guo:2017sjt,Cao:2020ibk}.

MC simulated data samples produced with a {\sc geant4}-based~\cite{GEANT4:2002zbu} software package, which includes the geometric description of the BESIII detector~\cite{BESIII:2009fln} and the detector response, are used to determine detection efficiencies, estimate backgrounds, and train the deep learning model. The simulation models the beam energy spread and initial state radiation (ISR) in the $e^+e^-$ annihilations with the generator {\sc kkmc}~\cite{Jadach:2000ir,Jadach:1999vf}. The inclusive MC sample includes the production of $\Lambda_c^+\bar{\Lambda}_c^-$ pairs, open-charm meson, the ISR production of vector charmonium(-like) states, and the continuum processes incorporated in {\sc kkmc}~\cite{Jadach:2000ir,Jadach:1999vf}. All particle decays are modelled with {\sc evtgen}~\cite{Lange:2001uf,Ping:2008zz} using BFs either taken from the Particle Data Group (PDG)~\cite{ParticleDataGroup:2024cfk}, when available, or otherwise estimated with {\sc lundcharm}~\cite{Yang:2014vra}. Final state radiation from charged final state particles is incorporated using the {\sc photos} package~\cite{Barberio:1990ms}. The $e^+e^-\to\Lc\ALc$ signal MC sample is generated with $\Lambda_c^+$ decaying via the signal process $\Lc\to X\pi^+\eta\,(X=n,\,\Lambda,\,\Sigma^{0})$ while the other $\ALc$ decays through 11 single-tag (ST) modes, as described below. The signal processes $\Lc\to n\pi^+\eta$ and $\Sigma^0\pi^+\eta$ are simulated with a uniformly distributed phase space (PHSP) model, while the decay $\Lc\to\Lambda\pi^+\eta$ is simulated based on the PWA model~\cite{BESIII:2024mbf}. For two-body $\ALc$ ST modes, the angular distributions are described by the transverse polarisation and decay asymmetry parameters of $\ALc$ and its daughter baryons~\cite{Chen:2019hqi}. For three-body and four-body $\ALc$ ST modes, the intermediate states are modelled according to individual internal PWA models.

\section{\boldmath Branching fraction measurement of \texorpdfstring{$\Lc\to n\pi^+\eta$}{Lambdac to n pi+ eta}}\label{sec:BF}

\subsection{Analysis method}\label{sec:method}
\hspace{1.5em}
Taking advantage of $\Lc\ALc$ pair production in $e^+e^-$ annihilation at centre-of-mass energies between $\sqrt{s}=4.600$ and $4.843~\gev$, we employ a double-tag (DT) method~\cite{MARK-III:1985hbd} to constrain the kinematics of $\Lcnpieta$. The $\ALc$ baryon is first reconstructed in any of the following 11 exclusive hadronic decay modes: $\ALc\to \bar{p}K_S^0$, $\bar{p} K^+ \pi^-$, $\bar{p} K_S^0 \pi^0$, $\bar{p} K_S^0 \pi^- \pi^+$, $\bar{p} K^+ \pi^- \pi^0$, $\bar{\Lambda} \pi^-$, $\bar{\Lambda} \pi^- \pi^0$, $\bar{\Lambda} \pi^- \pi^+ \pi^-$, $\bar{\Sigma}^0 \pi^-$, $\bar{\Sigma}^- \pi^0$, and $\bar{\Sigma}^- \pi^- \pi^+$, referred to as the ST candidate. In the presence of an ST $\ALc$ baryon, the signal decay of $\Lc$ is then searched for in its recoil system, with selected events referred to as DT events. The BF of the signal decay is determined by

\begin{equation}
\mathcal{B}_{\rm{sig}}=\frac{\sum_{i,j}N^{i,j}_{\rm{DT}}}{\sum_{i,j}\left(\frac{N^{i,j}_{\rm{ST}}}{\epsilon_{\rm{ST}}^{i,j}}\cdot\epsilon_{\rm{DT}}^{i,j}\right)\cdot\mathcal{B}_{\mathrm{inter}}}
=\frac{N_{\rm{DT}}}{N_{\mathrm{ST}}\cdot\epsilon_{\mathrm{sig}}\cdot\mathcal{B}_{\mathrm{inter}}},
\label{eq:sigbf}
\end{equation}
where $N_{\rm{ST}}^{i,j}$, $N_{\rm{DT}}^{i,j}$, $\epsilon_{\rm{ST}}^{i,j}$, and $\epsilon_{\rm{DT}}^{i,j}$ are the ST yield, DT yield, ST efficiency, and DT efficiency for tag mode $i$ at energy point $j$, respectively. $\mathcal{B}_{\mathrm{inter}}$ is the BF of intermediate states taken from the PDG~\cite{ParticleDataGroup:2024cfk}. $N_{\mathrm{ST(DT)}} = \sum_{i,j} N_{\mathrm{ST(DT)}}^{i,j}$ is the total ST (DT) yield summed over all tag modes and energy points. $\epsilon_{\mathrm{sig}} = \sum_{i,j}\left(\frac{N^{i,j}_{\rm{ST}}}{\epsilon_{\rm{ST}}^{i,j}}\cdot\epsilon_{\rm{DT}}^{i,j}\right) / N_{\mathrm{ST}}$ is the effective detection efficiency of DT events in the presence of ST $\ALc$.

In selection of the signal decay, the neutron is not directly reconstructed but represented as a recoil against the tagged $\ALc$, the signal $\pi^+$ and $\eta$. Without imposing prior constraints on the recoiling information, the decays $\Lclmdpieta$ and $\Lcsgmpieta$ can also appear with substantial yields~\cite{ParticleDataGroup:2024cfk}. This approach enables the simultaneous extraction of signals for the decays $\Lcxpieta$ within this analysis. 

To observe clear signals against background events, a deep-learning-based signal identification method is eventually applied to the DT data sample. Considering the potential bias it may introduce to the resulting BFs, we opt to report the relative BF of $\Lcnpieta$ with respect to $\Lclmdpieta$, which is expected to cancel the leading bias since these two decays share similar topologies and final states~\cite{BESIII:2024cbr}. Any residual bias is further treated as a source of systematic uncertainty, quantified using $\Lcsgmpieta$ as a control channel.

\subsection{Primary event selection}\label{sec:eventselection}
\hspace{1.5em}
The event selection criteria for the ST $\ALc$ candidate exactly follow the previous BESIII analysis~\cite{BESIII:2026qbp}, where detailed ST yields and detection efficiencies for various ST modes at each energy point can be found. The signal candidates for $\Lcxpieta$ are selected using the remaining charged tracks and showers in the presence of the ST $\ALc$ baryon, as detailed below.

Charged tracks detected in the MDC are required to be within a polar angle ($\theta$) range of $|\rm{cos\theta}|<0.93$, where $\theta$ is defined with respect to the $z$-axis, which is the symmetry axis of the MDC. The distance of closest approach to the interaction point (IP) must be less than 10\,cm along the $z$-axis, $|V_{z}|$, and less than 1\,cm in the transverse plane, $|V_{xy}|$. After removing the tracks used in the ST $\bar{\Lambda}_c^-$ candidate, the recoil side is required to contain exactly one loosely selected charged track for the $\eta\to\gamma\gamma$ mode ($|V_{z}|<20\,\mathrm{cm}$, $|\rm{cos\theta}|<0.93$), or exactly three loosely selected charged tracks for the $\eta\to\pi^+\pi^-\pi^0$ mode. The net charge of these recoil-side tracks is required to be opposite to that of the ST $\bar{\Lambda}_c^-$ candidate. The one-track requirement corresponds to the signal $\pi^+$ in the $\eta\to\gamma\gamma$ mode, while the three-track requirement corresponds to the signal $\pi^+$ and the $\pi^+\pi^-$ pair from $\eta\to\pi^+\pi^-\pi^0$. With this requirement, the $\Lambda\pi^+\eta$ and $\Sigma^0\pi^+\eta$ modes are also retained when $\Lambda$ decays dominantly through $\Lambda\to n\pi^0$ and $\Sigma^0$ through $\Sigma^0\to\Lambda[\to n\pi^0]\gamma$.

Particle identification~(PID) for charged tracks combines measurements of the specific ionization energy loss in the MDC~(d$E$/d$x$) and the flight time in the TOF to form likelihoods $\mathcal{L}(h)~(h=K,\pi)$ for each hadron $h$ hypothesis. Tracks are identified as pions by comparing the likelihoods for the kaon and pion hypotheses, $\mathcal{L}(\pi)>\mathcal{L}(K)$.

Photon candidates are identified using isolated showers in the EMC. The deposited energy of each shower must be more than 25~MeV in the barrel region ($|\cos \theta|< 0.80$) and more than 50~MeV in the end cap region ($0.86 <|\cos \theta|< 0.92$). To exclude showers that originate from charged tracks, the angle subtended by the EMC shower and the position of the closest charged track at the EMC must be greater than 10 degrees as measured from the IP. To suppress electronic noise and showers unrelated to the event, the difference between the EMC time and the event start time must be within [0, 700]\,ns. At least two photon candidates are required in each event, as they are needed to reconstruct either $\eta\to\gamma\gamma$ or the $\pi^0\to\gamma\gamma$ decay in $\eta\to\pi^+\pi^-\pi^0$. Additional photons are allowed and are considered in the neutral-meson candidate selection described below.

The $\eta$ meson is reconstructed via its two dominant decay modes $\etagg$ and $\etatpi[\to\gamma\gamma]$. For $\etagg\;(\pi^0\to\gamma\gamma)$, the invariant mass of two photons, $M_{\gamma\gamma}$, is required to be in the range [0.505, 0.575]$\,\mathrm{GeV}/c^2$ ([0.115, 0.150]$\,\mathrm{GeV}/c^2$). To improve the momentum resolution, a one-constraint (1C) kinematic fit is performed by constraining $M_{\gamma\gamma}$ to the known $\eta\;(\pi^0)$ mass~\cite{ParticleDataGroup:2024cfk}, and the fit $\chi^2$ must be less than 20 (200). For $\etatpi$, the invariant mass of $\pi^+\pi^-\pi^0$, $M_{\pi^+\pi^-\pi^0}$, is required to be in the range [0.505, 0.575]$\,\mathrm{GeV}/c^2$. A 1C kinematic fit is performed by constraining $M_{\pi^+\pi^-\pi^0}$ to the known $\eta$ mass~\cite{ParticleDataGroup:2024cfk}, and the fit $\chi^2$ must be less than 20. The updated momenta from the kinematic fit are used in further analysis. The average $\eta$ candidate multiplicities are approximately 1.3 for both $\eta$ decay modes, according to signal MC simulations. If multiple $\etagg$ or $\etatpi$ candidates exist in an event, the one with minimum $\chi^2$ value is retained for each $\eta$ decay mode for further analysis.

To derive information on the missing $X(n,\,\Lambda,\,\Sigma^{0})$, a kinematic observable $M_{\rm miss}$ is defined as $M_{\rm miss}\equiv\sqrt{E^2_{\rm miss}/c^4-\left|\vec{p}_{\rm miss}\right|^2/c^2}$, where $E_{\rm miss}$ and $\vec{p}_{\rm miss}$ are the total energy and momentum of all missing particles in the event, respectively. $E_{\rm miss}$ is calculated as $E_{\rm miss}=E_{\rm beam}-E_{\pi^+}-E_{\eta}$, where $E_{\rm beam}$ is the beam energy, and $E_{\pi^+}$ and $E_{\eta}$ are the energies of the signal pion and $\eta$ meson, respectively. The constrained $\Lc$ momentum is used to calculate $\vec{p}_{\rm miss}$ as $\vec{p}_{\rm miss}=\vec{p}_{\Lc}-\vec{p}_{\pi^+}-\vec{p}_{\eta}$, where $\vec{p}_{\Lc}$, $\vec{p}_{\pi^+}$, and $\vec{p}_{\eta}$ are the momenta of the signal $\Lc$ baryon, pion, and $\eta$ meson, respectively. $\vec{p}_{\Lc}$ is further given by $\vec{p}_{\Lc}=-\frac{\vec{p}_{\ALc}}{\left|\vec{p}_{\ALc}\right|}\sqrt{E^2_{\rm beam}/c^2-m^2_{\Lc}c^2}$, where $\vec{p}_{\ALc}$ is the momentum of the ST $\ALc$ baryon and $m_{\Lc}$ is the nominal mass~\cite{ParticleDataGroup:2024cfk}.

With the help of a generic event type analysis tool, TopoAna~\cite{Zhou:2020ksj}, some primary background components are identified in the inclusive MC sample and effectively reduced. For the $\etagg$ decay channel, to suppress $\Lc\to\Sigma^+[\to n\pi^+]\eta$ peaking background events, we apply a veto on the recoil mass spectrum of $\eta$, $M_{\mathrm{recoil}}(\eta)\not\in (1.13,1.25)\gev/c^2$. For the $\etatpi$ decay channel, to suppress $\Lc\to\Sigma^+[\to n\pi^+]\eta$ background events, we apply an analogous veto on the recoil mass spectrum of $\eta$, $M_{\mathrm{recoil}}(\eta)\not\in (1.16,1.22)\gev/c^2$. To suppress $\Lc\to\Lambda[\to n\pi^0]\pi^+\pi^-\pi^+$ background events, we apply a veto on the recoil mass spectrum of $\pi^+\pi^-\pi^+$, $M_{\mathrm{recoil}}(\pi^+\pi^-\pi^+)\not\in (1.10,1.13)\gev/c^2$. The veto regions correspond to approximately $\pm3\sigma$ of the background resolutions. After applying these vetoes, the resulting $M_{\mathrm{miss}}$ distributions for the $\etagg$ and $\etatpi$ channels are shown in Fig.~\ref{fig:cutbased_mm_vetocut}, where the high background level impedes observation of the $\Lcnpieta$ signals with adequate statistical significance. These backgrounds comprise a great variety of hadronic final states involving $\Lambda^+_c$ as an intermediate state or not, thereby requiring further signal identification treatments as elaborated below.

\begin{figure}[htbp]
\centering
\subfigure[$\etagg$]{\includegraphics[width=0.48\textwidth]{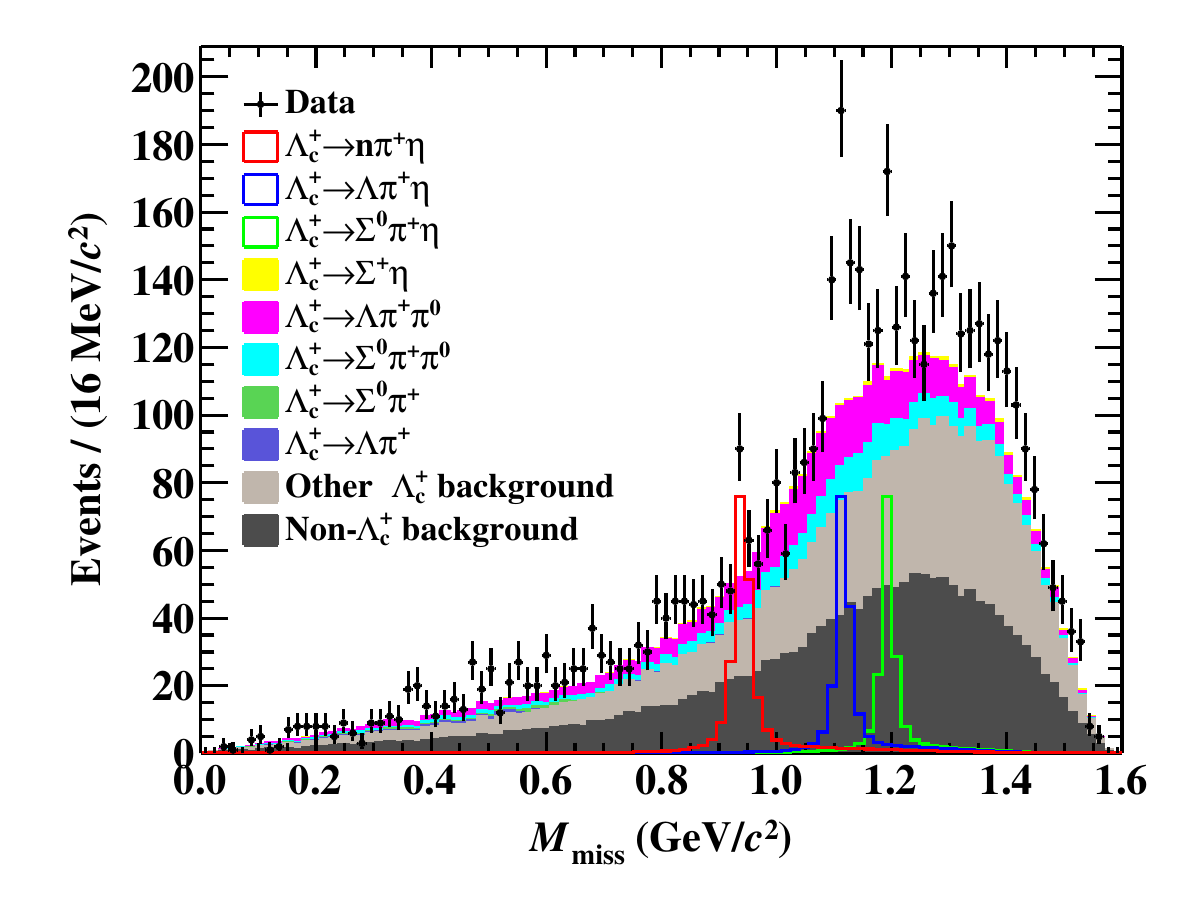}}
\subfigure[$\etatpi$]{\includegraphics[width=0.48\textwidth]{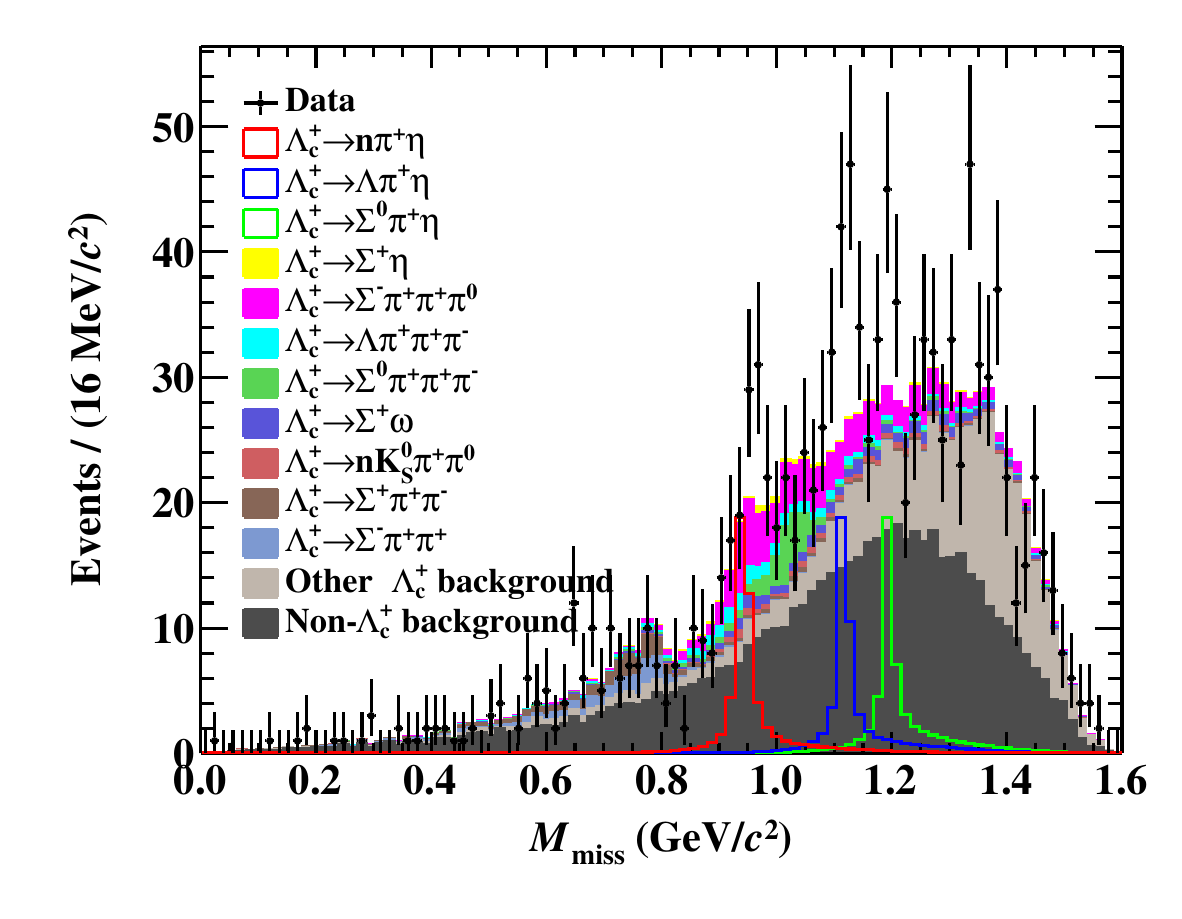}}
\caption{The $\mm$ distributions for the $\etagg$ and $\etatpi$ channels after veto cuts. The points with error bars represent data. The histograms with filled colors represent different background components in the inclusive MC sample. The histograms with blank fillings indicate arbitrarily scaled signal shapes.}
\label{fig:cutbased_mm_vetocut}
\end{figure}

\subsection{Deep-learning-based signal identification}\label{sec:gnn}
\hspace{1.5em}
The main challenge in distinguishing signals from various backgrounds lies in extracting distinctive features from the data. These can include the kinematics, identification confidence, and reconstruction quality for each final-state particle, as well as the overall topology and dynamics of the $e^+e^-$ decay chain. Conventional identification methods characterise such distinctions via hand-crafted variables derived from practical understanding of the involved physics, feeding them into multivariate analysis tools such as boosted decision trees (BDTs)~\cite{Hocker:2007ht}. In contrast, deep learning methods aim to process minimally refined data at the level of fundamental detector responses to the penetration of final state particles. The DNN is expected to automatically explore underlying correlations and recognise distinctive information beyond prior knowledge.

The dataset for training the DNN is constructed by randomly shuffling signal and background events sourced from the relevant MC samples. The events are categorised into three groups with equal statistics: the $\Lcxpieta$ signal, the $\Lc$ background, and the non-$\Lc$ background. The total training dataset contains approximately 0.7 million events, with 80\% used for training and the remaining 20\% reserved for validation.

The input information to the DNN from an event includes all charged tracks reconstructed in the MDC and isolated showers clustered in the EMC. These particles are organised as a {\it point cloud}~\cite{Qu:2019gqs} structure, i.e., an unordered and variable-sized set of points in a high-dimensional feature space. For each charged track, the features comprise the azimuthal and polar angles in the centre-of-mass frame, the charge, the magnitude of momentum, and parameters characterising its helical trajectory in the MDC. Additionally, low-level measurements from the MDC, TOF, and EMC are incorporated as implicit information for particle identification and reconstruction quality. For each shower, the features consist of the azimuthal and polar angles, the energy deposition, the count of fired crystals, the time measurement, and parameters describing its expansion scope among nearby crystals. A full list of input features is provided in Ref.~\cite{BESIII:2025biy}, and these features are found to be overall consistent between data and MC simulation.

The architecture of the DNN largely inherits from the Particle Transformer (ParT)~\cite{qu2022particle}, with adaptations tailored for the BESIII experiment as detailed in Ref.~\cite{BESIII:2025biy}. The model hyperparameters have been optimised to maximise DNN performance while preventing overfitting. In addition, a machine learning technique called model ensemble is employed. With randomness factors incorporated in training such as network initialization, batch processing sequence, and dropout~\cite{Srivastava:2014kpo} mechanism, a total of 50 DNNs are trained in parallel. The outputs of these DNNs are averaged for each event during inference, creating an ensemble DNN that offers better robustness and generalization than any single DNN model.

The DNN output assigns three scores in $[0,\,1]$ to each event, reflecting the probabilities of the event belonging to the $\Lcxpieta$ signal, the $\Lc$ background, and the non-$\Lc$ background categories. Since the sum of the three scores always equals one, only two of them are independent. As the final step in event selection, we require the score for $\Lc$ background to be less than 0.05 and the score for non-$\Lc$ background to be less than 0.1. These cut values are optimised by maximizing the figure of merit $\frac{S}{\sqrt{S+B}}$ for $\Lcnpieta$, where $S(B)$ denotes the expected signal (background) yield from an MC study with data-matched luminosities.

Figure~\ref{fig:Fit} illustrates the $M_{\rm miss}$ distributions after deep learning implementation. The backgrounds are significantly reduced in both channels, to approximately 1/20 of their original levels; this enables clear observation of signals for $\Lcxpieta$. The effective detection efficiencies $\epsilon_{\rm sig}$ for each decay mode are listed in Table~\ref{tab:eff}, where the uncertainties are statistical only.

\begin{table}[ht]
\centering
\caption{Summary of effective detection efficiencies after DNN implementation.}
\label{tab:eff}
\begin{tabular}{lc}
\hline \hline
Decay mode                  &  $\epsilon_{\mathrm{sig}}$ (\%)  \\
\hline
$\Lcnpieta,\ \etagg$      & $31.49\pm0.05$ \\
$\Lclmdpieta,\ \etagg$      & $13.43\pm0.03$ \\
$\Lcsgmpieta,\ \etagg$      & $12.21\pm0.03$ \\
$\Lcnpieta,\ \etatpi$      & $5.90\pm0.02$ \\
$\Lclmdpieta,\ \etatpi$      & $3.83\pm0.02$ \\
$\Lcsgmpieta,\ \etatpi$      & $3.83\pm0.02$ \\
\hline \hline
\end{tabular}
\end{table}

\subsection{Signal extraction}\label{sec:BFfit}
\hspace{1.5em}
To obtain the DT yield $N_{\rm DT}$ in data, unbinned maximum likelihood fits are performed on the $\mm$ spectra. These fits are conducted simultaneously on both $\etagg$ and $\etatpi$ channels, sharing the absolute BFs of $\Lcxpieta$ signals. In each fit, the probability distribution function (PDF) includes five components: the $\Lcnpieta$ signal, the $\Lclmdpieta$ signal, the $\Lcsgmpieta$ signal, the $\Lambda^+_c$ background, and the non-$\Lambda^+_c$ background. The shapes for these components are taken from the corresponding MC simulations, and those of signals are further convolved with a Gaussian function representing the resolution difference. The parameters of the Gaussian function are shared among the three signals but separated between the $\etagg$ and $\etatpi$ channels. The ratio of $\Lambda^+_c$ and non-$\Lambda^+_c$ background yields is fixed according to MC simulation.

Figure~\ref{fig:Fit} shows the fit results. The fits obtain $59.7\pm9.9$ signal events for $\Lcnpieta$, $176.1\pm19.3$ for $\Lclmdpieta$, and $71.2\pm13.0$ for $\Lcsgmpieta$, with all uncertainties being statistical only. Based on the likelihood and degrees of freedom variations when canceling one signal component in the PDF, the statistical significance of $\Lcnpieta$ is calculated to be $9.5\sigma$, yielding the first experimental observation of $\Lcnpieta$. According to Eq.~(\ref{eq:sigbf}) and the discussion in Section~\ref{sec:method}, we calculate the relative BF of $\Lcnpieta$ with respect to $\Lclmdpieta$ as $0.155\pm0.031_{\rm stat.}$. For validation checks on the analysis strategy, the absolute BFs for $\Lclmdpieta$ and $\Lcsgmpieta$ are extracted to be $(1.87\pm0.21_{\rm stat.})\times10^{-2}$ and $(8.2\pm1.5_{\rm stat.})\times10^{-3}$, respectively, consistent with previous measurements~\cite{ParticleDataGroup:2024cfk}.

\begin{figure}[htbp]
\centering
\subfigure[$\etagg$]{\includegraphics[width=0.48\textwidth]{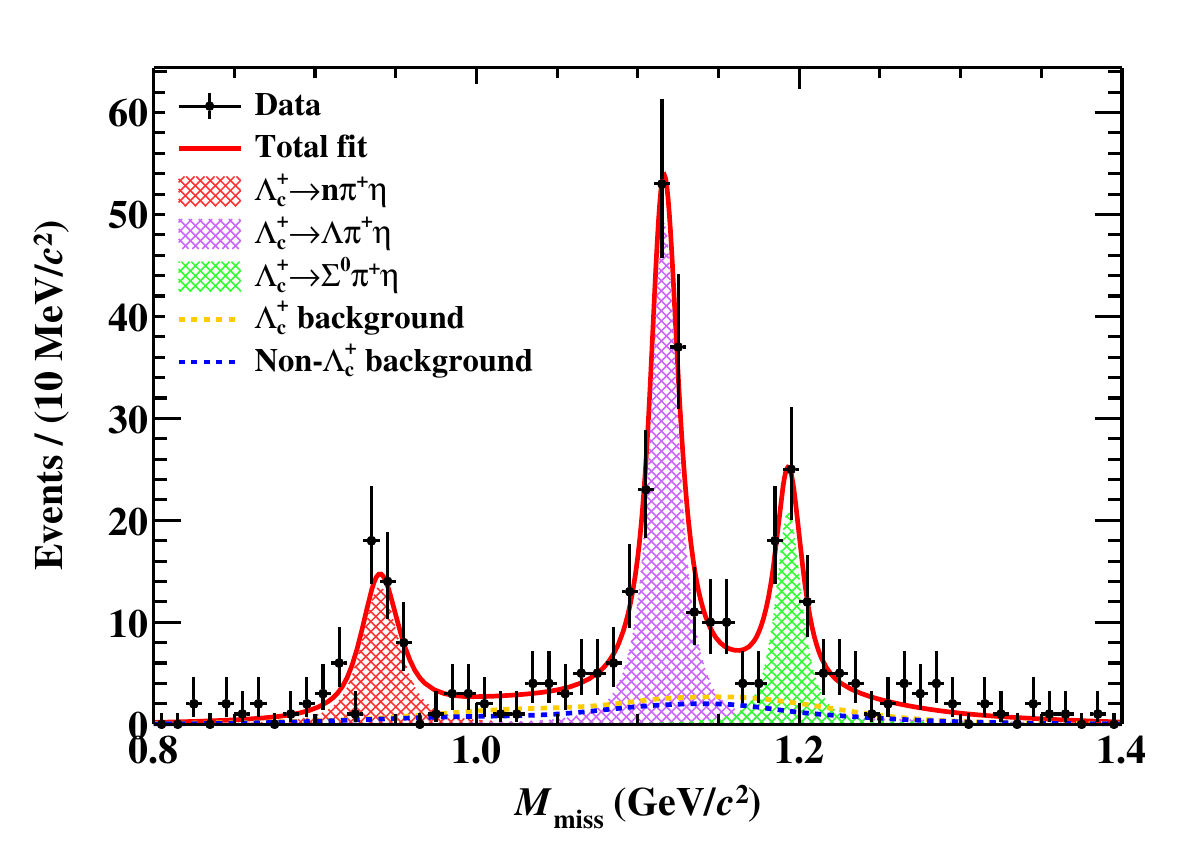}}
\subfigure[$\etatpi$]{\includegraphics[width=0.48\textwidth]{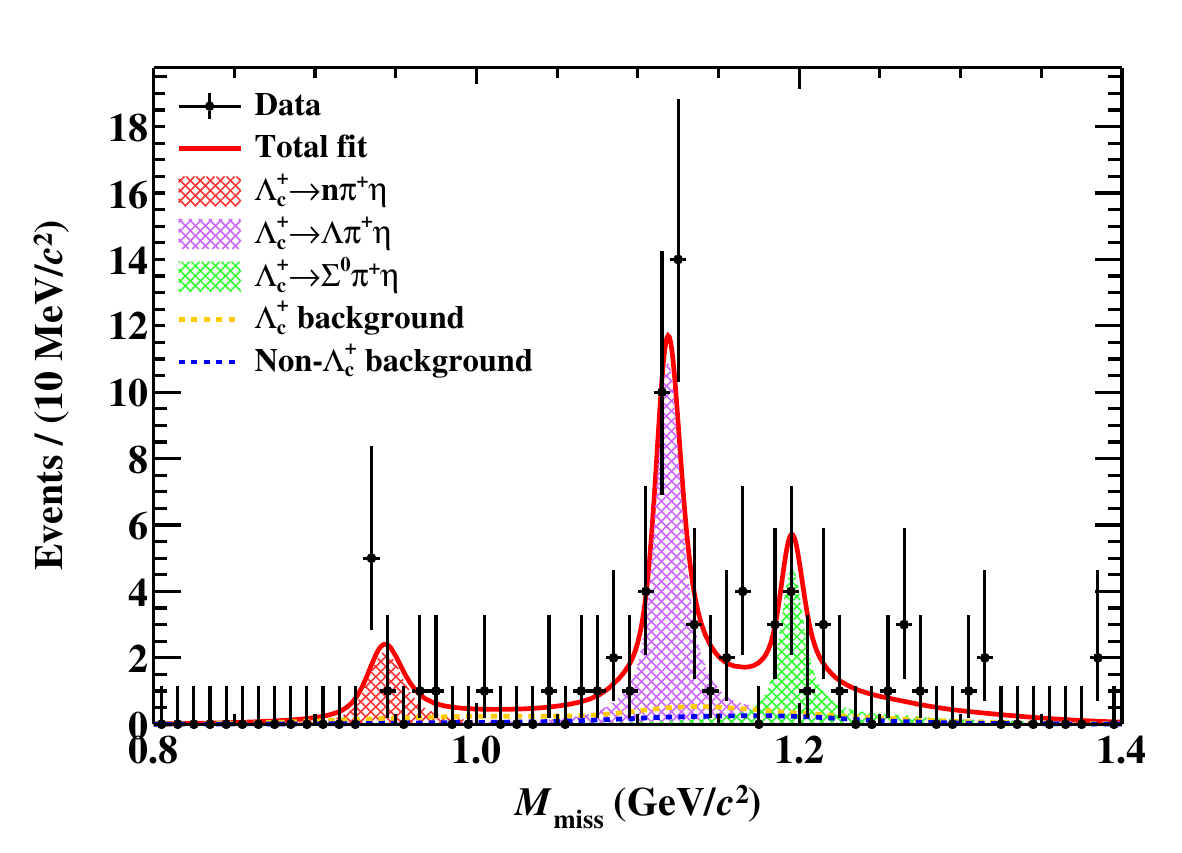}}
\caption{Post-fit $\mm$ spectra of data samples. The points with error bars represent data. The red solid lines represent the total fit results. The dashed areas represent signal components and the dashed lines represent background components.}
\label{fig:Fit}
\end{figure}

\subsection{Systematic uncertainties}\label{sec:BFsys}
\hspace{1.5em}
With the DT technique and the relative BF measurement strategy, many uncertainty sources for the BF ratio $\mathcal{B}(\Lcnpieta)/\mathcal{B}(\Lclmdpieta)$ are canceled, including the ST analysis procedure, the reconstruction for signal pion and $\eta$ meson, and the input BFs for intermediate decays. The remaining sources of systematic uncertainty are summarised in Table~\ref{tab:sys_err}.

\begin{table}[ht]
\centering
\caption{Systematic uncertainties for the measurement of the BF ratio $\mathcal{B}(\Lcnpieta)/\mathcal{B}(\Lclmdpieta)$.}
\label{tab:sys_err}
\begin{tabular}{lc}
\hline \hline
Sources                      & Size (\%)  \\
\hline
MC model                     & 5.7 \\
Peaking background vetoes    & 1.5 \\
Deep learning selections      & 4.9 \\
Simultaneous fit             & 0.9 \\
\hline
Total                        & 7.7 \\
\hline \hline
\end{tabular}
\end{table}

\begin{itemize}
    \item {\it MC model.} The uncertainty from MC modelling for $\Lcnpieta$ is estimated using a BDT-based reweighting method~\cite{Rogozhnikov:2016bdp}, where the simulated $M_{\mathrm{recoil}}(\pi^{+})$, $M_{\mathrm{recoil}}(\eta)$, and $M_{\pi^{+}\eta}$ distributions are tuned to approximate the data distributions. The same uncertainty source for $\Lclmdpieta$ is estimated by varying amplitude model parameters according to their error matrices. By inputting the new detection efficiencies of $\Lcnpieta$ and $\Lclmdpieta$ into the simultaneous fit, the averaged difference in the resulting BF ratio of 5.7\% is taken as the corresponding systematic uncertainty.

    \item {\it Peaking background vetoes.} The uncertainty due to veto regions for $M_{\mathrm{recoil}}(\eta)$ and $M_{\mathrm{recoil}}(\pi^+\pi^-\pi^+)$ is calculated by changing the lower and upper bounds of these regions by $\pm5\,\mathrm{MeV}/c^2$ arbitrarily. The averaged shift of the resulting BF ratio is calculated to be 1.5\% and is assigned as a source of systematic uncertainty.

    \item {\it Deep learning selections.} The uncertainties related to the DNN selections are evaluated following the method in Refs.~\cite{BESIII:2024cbr,BESIII:2025biy}, which considers two major sources: {\it model uncertainty} and {\it domain shift}. The uncertainty due to the model arises from our limited knowledge of the best model and is quantified using alternative DNNs trained with different hyperparameters. The shift in the resulting BF ratio of 0.8\% is assigned as the corresponding uncertainty value. On the other hand, domain shift describes the mismatch between datasets for training and inference, reflecting potential data-MC inconsistency in this study. We assume such uncertainty to be mostly canceled in the relative BF due to the similarity of final states in $\Lcnpieta$ and $\Lclmdpieta$. This assumption is validated using the control channel $\Lcsgmpieta$, whose relative BF with respect to $\Lclmdpieta$ is found to be stable when the DNN selection cut values are scanned in steps of 0.1. The full BF extraction procedure is repeated at each scan point, and the largest deviation from the nominal result, 4.8\%, is assigned as the residual uncertainty.
    
    \item {\it Simultaneous fit.}  The systematic uncertainty associated with the simultaneous fit is mainly due to the modelling of signal and background components. A bootstrap resampling method~\cite{chernick2011bootstrap,BESIII:2020kzc} is employed to quantify the uncertainty, resulting in an assignment of 0.9\%. The uncertainty due to the signal shape is evaluated by varying the parameters of the Gaussian function. The uncertainty due to the $\Lc$ and non-$\Lc$ background shapes is evaluated by replacing the MC simulated shapes with 3rd order Chebyshev polynomials. The uncertainty due to the $\Lc$ and non-$\Lc$ background yields is evaluated by varying the fixed ratio arbitrarily.
\end{itemize}

Assuming that all sources are independent, the total systematic uncertainty is determined to be 7.7\% by adding the above sources in quadrature.

\section{\boldmath Search for \texorpdfstring{$\Lambda_c^+\to n a_0(980)^+$}{Lambdac to n a0(980)+}}\label{sec:fita0}

\subsection{Fit model and results}\label{sec:a0_fit}
\hspace{1.5em}
After the DNN selections, the intermediate process $\Lambda_c^+\to n a_0(980)^+, a_0(980)^+\to\pi^+\eta$ is searched for in the signal region $\mm\in(0.9,\ 1.0)\ \mathrm{GeV}/c^2$. An unbinned extended maximum likelihood fit is performed on the $\pi^+\eta$ invariant mass distribution.

The signal PDF is described by the incoherent superposition of an $a_0(980)^+$ Flatt\'{e} line shape and a non-resonant PHSP distribution. For $a_0(980)^+$, the two coupled-channel Flatt\'{e} model~\cite{Flatte:1976xu} is used as
\begin{equation}
    \mathcal{R}_{a_0(980)^+}(m) = \frac{1}{m_0^2-m^2-i(g_1\rho_{\eta\pi}(m)+g_2\rho_{K\bar{K}}(m))},
\end{equation}
where $m$ refers to the $\pi^+\eta$ invariant mass. The nominal mass $m_0=(0.990\pm0.001)\gev/c^2$, the coupling constant for the
$\eta\pi$ channel, $g_1=(0.341 \pm 0.004)\gev^2/c^4$, and the coupling ratio
$g_2/g_1=0.892\pm0.022$ between the $K\bar{K}$ and $\eta\pi$ channels are quoted
from Ref.~\cite{BESIII:2016tqo}. $\rho_{\eta\pi}$ and $\rho_{K\bar{K}}$ are the available phase spaces for the $\eta\pi$ channel and the $K\bar{K}$ channel, obtained from the corresponding decay momentum $q$: $\rho = 2q/m$. When the decay momentum falls below the threshold $(m < m_{K} + m_{\bar{K}})$, the momentum becomes imaginary and contributes to the real parts of the propagator. 

Since no significant $a_0(980)^+$ signal is observed, the present data sample does not allow a reliable determination of the relative phase between the $a_0(980)^+$ and non-resonant amplitudes. Therefore, no coherent interference term is introduced in the nominal fit. The non-resonant component is described by a PHSP distribution, and the relative strength of the $a_0(980)^+$ component with respect to the non-resonant component is parametrized by a non-negative factor $r$. The full decay amplitude is
\begin{equation}
    \mathcal{A}(m) = r\times\mathcal{R}_{a_0(980)^+}(m) + 1.
\end{equation}
The partial decay width can be written as
\begin{equation}
    \frac{\mathrm{d}\Gamma}{\mathrm{d}m}=
    |\mathcal{A}(m)|^2\cdot
    \frac{|\vec{P}_{n}(m)|}{m_{\Lambda_c^+}}\cdot
    \frac{|\vec{P}_{\pi^+}(m)|}{m}\cdot
    2m,
\end{equation}
where $\frac{|\vec{P}_{n}|}{m_{\Lambda_c^+}}\cdot\frac{|\vec{P}_{\pi^+}|}{m}\cdot 2m$ is the PHSP factor with respect to the invariant mass of the $\pi^+\eta$ system, with $|\vec{P}_{n}|$ and $|\vec{P}_{\pi^+}|$ calculated in the centre-of-mass frame of $\Lambda_c^+\to n a_0(980)^+$ and $a_0(980)^+\to \pi^+\eta$, respectively,
\begin{equation}
\begin{aligned}
    |\vec{P}_{n}(m)| &= \frac{\lambda^{1/2}(m_{\Lambda_c^+}^2,m_{n}^2,m^2)}{2m_{\Lambda_c^+}},\\
    |\vec{P}_{\pi^+}(m)| &= \frac{\lambda^{1/2}(m^2,m_{\pi^+}^2,m_{\eta}^2)}{2m},\\
    \lambda(x,y,z) &= x^2+y^2+z^2-2xy-2xz-2yz.
\end{aligned}
\end{equation}

By neglecting interference between $a_0(980)^+$ and the non-resonant component, the amplitude can be simplified to $|\mathcal{A}(m)|^2 = r^2\times|\mathcal{R}_{a_0(980)^+}(m)|^2 + 1$. The signal PDF in the fit is described by
\begin{equation}
    \mathcal{PDF}_{\mathrm{sig}}(m) = [\epsilon(m) \times \frac{\mathrm{d}\Gamma}{\mathrm{d}m}]\otimes \mathcal{G}(\mu,\sigma),
\end{equation}
where $\epsilon(m)$ is the mass-dependent efficiency studied from signal MC simulation, and $\mathcal{G}(\mu,\sigma)$ is a Gaussian function accounting for detector resolution. The mean and resolution of the Gaussian function are fixed according to the study of the control sample $\Lambda_c^+\to\Lambda\pi^+\eta$. The total signal yield is constrained by a Gaussian function according to the fit result of the BF measurement, while the relative strength between $a_0(980)^+$ and the non-resonant component is allowed to float.

The background components consist of four parts: $\Lambda\pi^+\eta$, $\Sigma^0\pi^+\eta$, other $\Lambda_c^+$ background, and non-$\Lambda_c^+$ background. The shapes of these components are derived from MC simulations. The strengths of each background component are constrained by Gaussian functions according to the fit result of the BF measurement.

The fit result is shown in Fig.~\ref{fig:na980fit}, which gives $r=0.0\pm0.1$. Since no significant signal is observed, an upper limit on $\mathcal{B}(\Lambda_c^+\to na_0(980)^+)\times\mathcal{B}(a_0(980)^+\to\pi^+\eta)$ is set after accounting for all systematic uncertainties, as described below.

\begin{figure}[htbp]
\centering
\includegraphics[width=0.55\textwidth]{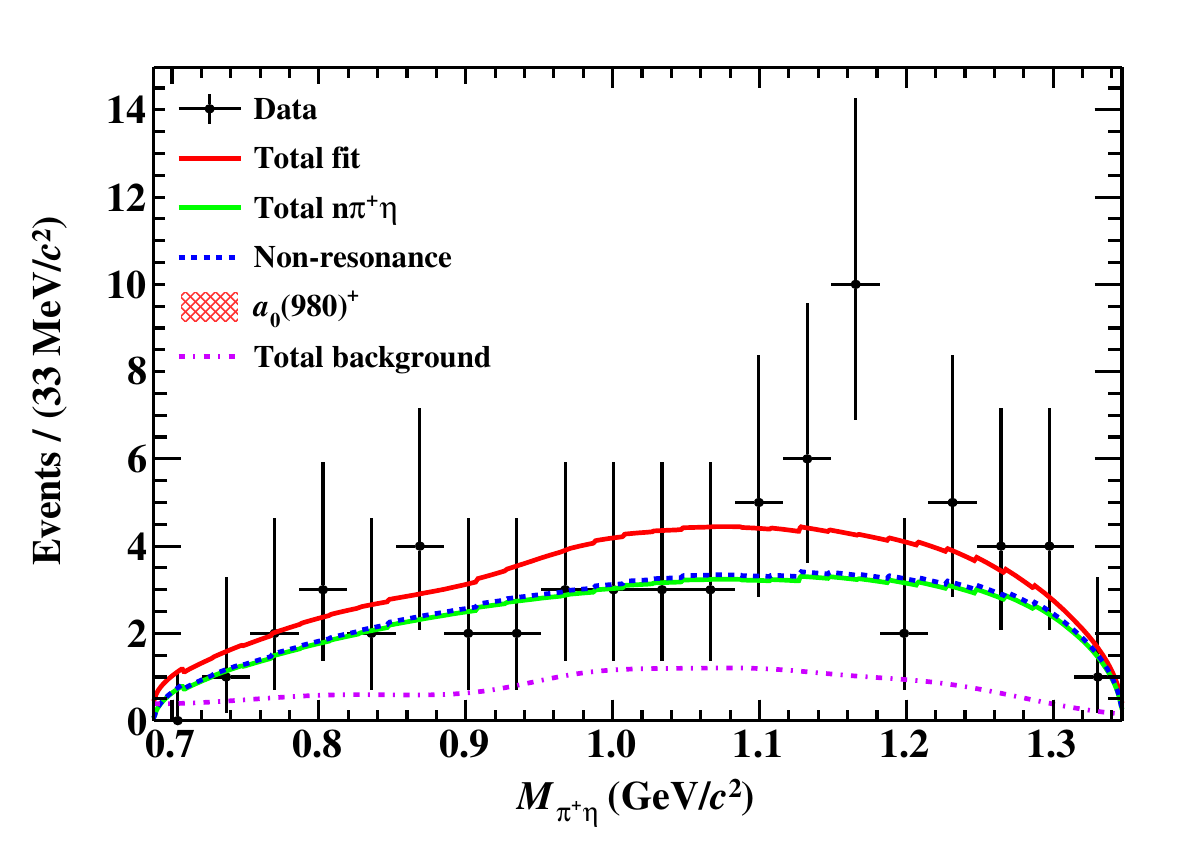}
\caption{Fit to the $M_{\pi^+\eta}$ distribution. The black points with error bars represent data. The red line represents the fit result. The green line represents the total $n\pi^+\eta$ signal. The blue dashed line represents the non-resonant $\pi^+\eta$ signal. The red dashed area represents the $na_0(980)^+$ signal, and the violet dash-dotted line is the total background including $\Lambda\pi^+\eta$, $\Sigma^0\pi^+\eta$, other $\Lambda_c^+$ background, and non-$\Lambda_c^+$ background.}
\label{fig:na980fit}
\end{figure}

\subsection{Systematic uncertainties}\label{sec:a0_syserr}
\hspace{1.5em}
The systematic uncertainties for $\mathcal{B}(\Lambda_c^+\to n a_0(980)^+)\times\mathcal{B}(a_0(980)^+\to\pi^+\eta)$ can be grouped into two categories: multiplicative uncertainties, which affect the BF of the three-body decay $\Lambda_c^+\to n\pi^+\eta$, and additive uncertainties, which impact the fraction of two-body decay to three-body decay. The multiplicative uncertainties have already been studied in Section~\ref{sec:BFsys}, while the additive uncertainties arise from signal shape, total yield constraint, background shape, and background magnitude, as discussed below.

\begin{itemize}
    \item {\it Signal shape.} We account for three potential sources affecting the signal shapes: efficiency curve, Gaussian resolution, and Flatt\'{e} parametrisation. For the efficiency curve, the binning is varied from $20\mev/c^2$ to a finer binning of $10\mev/c^2$ to assess bias from insufficient granularity. For Gaussian resolution, the mean and resolution of the Gaussian resolution function are varied by $\pm1\sigma$ according to their uncertainties. For Flatt\'{e} parametrisation, we vary all fixed parameters including mass and coupling constants by $\pm1\sigma$. Additionally, we test an alternative three coupled-channel parametrisation adding the $\pi^+\eta^\prime$ channel with parameters adjusted following Ref.~\cite{BESIII:2016tqo}.

    \item {\it Total yield constraint.} The systematic uncertainty is estimated by removing the Gaussian constraint on the signal yields and allowing them to float freely in the fit.

    \item {\it Background shape.} The background PDF is modified by reducing the kernel width to obtain a more detailed description of background distributions.

    \item {\it Background magnitude.} For each background component, we remove its Gaussian constraint individually to evaluate the corresponding systematic effect.

\end{itemize}

\subsection{Upper limit determination}\label{sec:a0_ul}
\hspace{1.5em}
The upper limit on $\mathcal{B}(\Lambda_c^+\to na_0(980)^+)\times\mathcal{B}(a_0(980)^+\to\pi^+\eta)$ is set using a Bayesian method~\cite{Liu:2015uha,Stenson:2006gwf}. We perform a series of maximum likelihood fits with the relative amplitude $r$ fixed to a scanning value. The corresponding maximum likelihood values are used to construct a discrete likelihood distribution, $\mathcal{L}(\mathcal{B})$. The additive systematic uncertainties are incorporated by varying the fit method as described in Section~\ref{sec:a0_syserr}, with the most conservative upper limit retained. The multiplicative uncertainties are accounted for by smearing the likelihood distribution according to the combined systematic uncertainties of $\mathcal{B}(\Lambda_c^+\to n\pi^+\eta)/\mathcal{B}(\Lambda_c^+\to\Lambda\pi^+\eta)$ and the uncertainty from $\mathcal{B}(\Lambda_c^+\to\Lambda\pi^+\eta)$~\cite{Blmdpieta}. By integrating the $\mathcal{L}(\mathcal{B})$ curve up to 90\% of the area for BF greater than zero, we set the upper limit on $\mathcal{B}(\Lambda_c^+\to na_0(980)^+)\times\mathcal{B}(a_0(980)^+\to\pi^+\eta)$ to be $8.4\times10^{-4}$ at 90\% confidence level, as shown in Fig.~\ref{fig:na980UL}.

%\begin{equation}
%    \mathcal{L}^{\prime}(\mathcal{B}_{\mathrm{2body}}) \propto \int_{0}^{1} \mathcal{L}(\mathcal{B}_{\mathrm{2body}}\cdot\frac{\hat{\mathcal{B}}_{\mathrm{3body}}}{\mathcal{B}_\mathrm{3body}})
%    e^{\frac{-(\mathcal{B}_\mathrm{3body}-\hat{\mathcal{B}}_{\mathrm{3body}})^2}{2\sigma_{\mathcal{B}_{\mathrm{3body}}}^2}}
%    d\mathcal{B}_{\mathrm{3body}},
%\end{equation}
%where $\hat{\mathcal{B}}_{\mathrm{3body}}$ is the nominal value of $\mathcal{B}(\Lambda_c^+\to n\pi^+\eta)$, and $\sigma_{\mathcal{B}_{\mathrm{3body}}}$ is the multiplicative uncertainty as discussed in Sec.~\ref{sec:BFsys} including the uncertainty from reference channel $\Lambda_c^+\to\Lambda\pi^+\eta$.

\begin{figure}[htbp]
\centering
\includegraphics[width=0.55\textwidth]{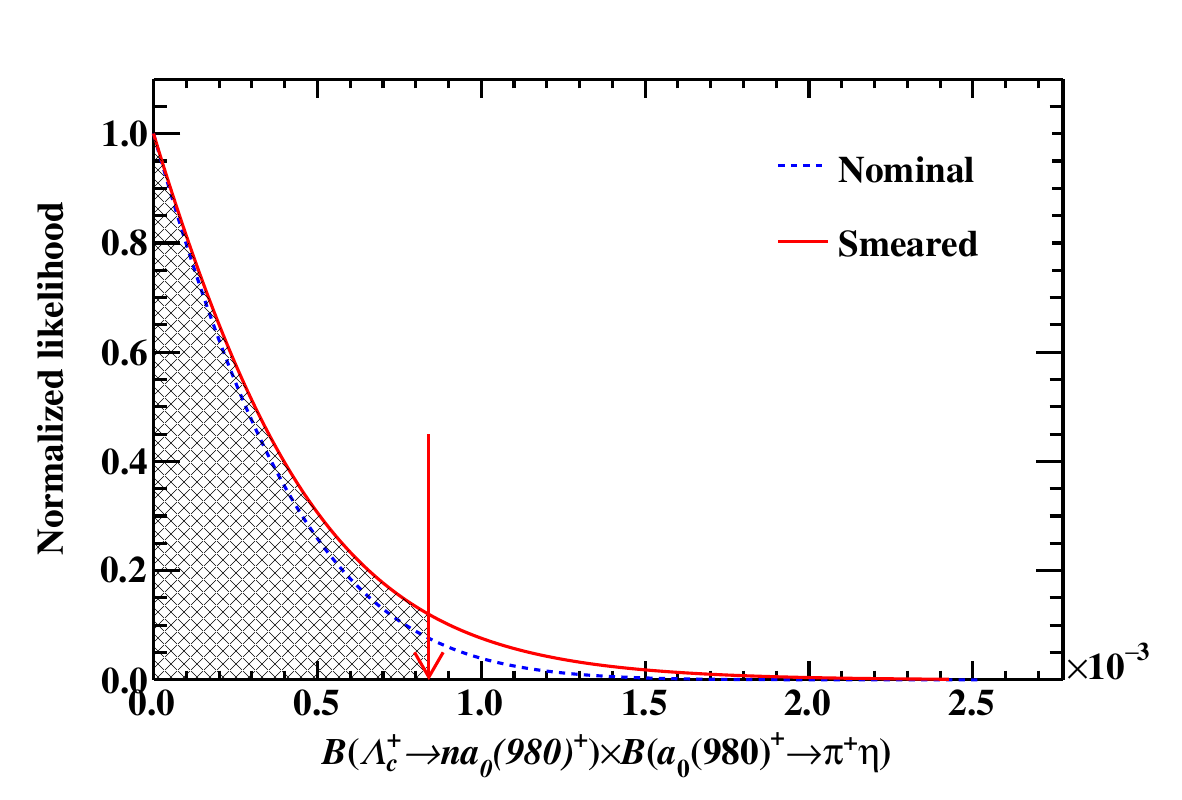}
\caption{The normalised likelihood distributions before and after incorporating systematic uncertainties. The blue dashed line represents the initial distribution and the red solid line represents the distribution with both additive and multiplicative systematic uncertainties considered. The red arrow indicates the final upper limit.}
\label{fig:na980UL}
\end{figure}

\section{Summary}\label{sec:summary}
\hspace{1.5em}
Based on 6.1~$\ifb$ of data collected at centre-of-mass energies between $\sqrt{s}=4.600$ and $4.843$~GeV with the BESIII detector, we observe the decay $\Lcnpieta$ for the first time with a statistical significance of $9.5\sigma$. Its relative BF with respect to $\Lclmdpieta$ is measured to be $\frac{\mathcal{B}(\Lcnpieta)}{\mathcal{B}(\Lclmdpieta)}=0.155\pm0.031_{\rm stat.}\pm0.012_{\rm syst.}$, where the first uncertainty is statistical and the second is systematic. Taking the world average of $\mathcal{B}(\Lclmdpieta)$~\cite{Blmdpieta} as reference, the absolute BF of $\Lcnpieta$ is calculated to be $\mathcal{B}(\Lcnpieta)=(2.94\pm0.59_{\rm stat.}\pm0.23_{\rm syst.}\pm0.13_{\rm ref.})\times10^{-3}$, where the last uncertainty is from the reference decay. Our measurement is consistent with the theoretical prediction in Ref.~\cite{Geng:2024sgq} within $2\sigma$, which considers contributions from both the color-antisymmetric $\bar{\textbf{6}}$ and color-symmetric $\textbf{15}$ components of the effective Hamiltonian.

For the decay $\Lambda^+_c\to na_{0}(980)^+$, no significant signal is observed in the $M_{\pi^+\eta}$ spectrum, and the upper limit on $\mathcal{B}(\Lambda_c^+\to na_0(980)^+)\times\mathcal{B}(a_0(980)^+\to\pi^+\eta)$ at 90\% confidence level is set to $8.4\times10^{-4}$, which is lower than the recent theoretical prediction using the IRA approach~\cite{Wang:2025uie}, but consistent with the earlier theoretical calculation using the pole model~\cite{Sharma:2009zze}. Dividing this upper limit by the measured BF of $\Lambda_c^+\to n\pi^+\eta$, we obtain $\mathcal{B}(\Lambda_c^+\to na_0(980)^+)\times\mathcal{B}(a_0(980)^+\to\pi^+\eta)/
\mathcal{B}(\Lambda_c^+\to n\pi^+\eta)
< 0.29$ at 90\% confidence level, which is close to the theoretical prediction in the chiral unitary approach~\cite{Li:2025gvo}. With the larger data sample to be collected at BESIII~\cite{BESIII:2020nme,Li:2021iwf,Li:2025nzx,Wang:2024wrm}, the sensitivity for this decay will be further improved to deepen the understanding of charmed baryon decays.

Technically, this success is largely due to the application of a Transformer-based deep learning technique, which is particularly effective at distinguishing signal from substantial background noise. Building on our previous research in charmed hadron decays~\cite{BESIII:2024mgg,BESIII:2024cbr,BESIII:2025biy}, and supported by comprehensive validation and uncertainty assessment tools, our results highlight the transformative potential and broad applicability of deep learning methods in BESIII data analysis.

\acknowledgments
\hspace{1.5em}
%% Saved at => 2025-09-26
%\textbf{Acknowledgement}

The BESIII Collaboration thanks the staff of BEPCII (https://cstr.cn/31109.02.BEPC) and the IHEP computing center for their strong support. This work is supported in part by National Key R\&D Program of China under Contracts Nos. 2023YFA1606000, 2023YFA1606704; National Natural Science Foundation of China (NSFC) under Contracts Nos. 11635010, 11935015, 11935016, 11935018, 12025502, 12035009, 12035013, 12061131003, 12192260, 12192261, 12192262, 12192263, 12192264, 12192265, 12221005, 12225509, 12235017, 12342502, 12361141819; the Chinese Academy of Sciences (CAS) Large-Scale Scientific Facility Program; CAS Project for Young Scientists in Basic Research No. YSBR-117; the Strategic Priority Research Program of Chinese Academy of Sciences under Contract No. XDA0480600; CAS under Contract No. YSBR-101; Postdoctoral Fellowship Program of China Postdoctoral Science Foundation under Grant No. GZC20261827; 100 Talents Program of CAS; The Institute of Nuclear and Particle Physics (INPAC) and Shanghai Key Laboratory for Particle Physics and Cosmology; ERC under Contract No. 758462; German Research Foundation DFG under Contract No. FOR5327; Istituto Nazionale di Fisica Nucleare, Italy; Knut and Alice Wallenberg Foundation under Contracts Nos. 2021.0174, 2021.0299, 2023.0315; Ministry of Development of Turkey under Contract No. DPT2006K-120470; National Research Foundation of Korea under Contract No. NRF-2022R1A2C1092335; National Science and Technology fund of Mongolia; Polish National Science Centre under Contract No. 2024/53/B/ST2/00975; STFC (United Kingdom); Swedish Research Council under Contract No. 2019.04595; U. S. Department of Energy under Contract No. DE-FG02-05ER41374.

\bibliographystyle{JHEP}
\bibliography{bibitem}

\newpage
\section*{The BESIII collaboration}
\addcontentsline{toc}{section}{The BESIII collaboration}
%% Saved at => 2025-09-26
M.~Ablikim$^{1}$\BESIIIorcid{0000-0002-3935-619X},
M.~N.~Achasov$^{4,b}$\BESIIIorcid{0000-0002-9400-8622},
P.~Adlarson$^{81}$\BESIIIorcid{0000-0001-6280-3851},
X.~C.~Ai$^{87}$\BESIIIorcid{0000-0003-3856-2415},
C.~S.~Akondi$^{31A,31B}$\BESIIIorcid{0000-0001-6303-5217},
R.~Aliberti$^{39}$\BESIIIorcid{0000-0003-3500-4012},
A.~Amoroso$^{80A,80C}$\BESIIIorcid{0000-0002-3095-8610},
Q.~An$^{77,64,\dagger}$,
Y.~H.~An$^{87}$\BESIIIorcid{0009-0008-3419-0849},
Y.~Bai$^{62}$\BESIIIorcid{0000-0001-6593-5665}
% ...
\\
\vspace{0.2cm}
(BESIII Collaboration)\\
\vspace{0.2cm} {\it
$^{1}$ Institute of High Energy Physics, Beijing 100049, People's Republic of China\\
% ...
$^{4}$ Budker Institute of Nuclear Physics SB RAS (BINP), Novosibirsk 630090, Russia\\
% ...
$^{31}$ INFN Sezione di Ferrara, (A)INFN Sezione di Ferrara, I-44122, Ferrara, Italy; (B)University of Ferrara, I-44122, Ferrara, Italy\\
% ...
$^{39}$ Johannes Gutenberg University of Mainz, Johann-Joachim-Becher-Weg 45, D-55099 Mainz, Germany\\
% ...
$^{62}$ Southeast University, Nanjing 211100, People's Republic of China\\
% ...
$^{64}$ State Key Laboratory of Particle Detection and Electronics, Beijing 100049, Hefei 230026, People's Republic of China\\
% ...
$^{77}$ University of Science and Technology of China, Hefei 230026, People's Republic of China\\
% ...
$^{80}$ University of Turin and INFN, (A)University of Turin, I-10125, Turin, Italy; (B)University of Eastern Piedmont, I-15121, Alessandria, Italy; (C)INFN, I-10125, Turin, Italy\\
$^{81}$ Uppsala University, Box 516, SE-75120 Uppsala, Sweden\\
% ...
$^{87}$ Zhengzhou University, Zhengzhou 450001, People's Republic of China\\

\vspace{0.2cm}
$^{\dagger}$ Deceased\\
% ...
$^{b}$ Also at the Novosibirsk State University, Novosibirsk, 630090, Russia\\
% ...

}
%% ends here %%

\end{document}